\documentclass[pre,twocolumn,superscriptaddress,nofootinbib]{revtex4-2}

\usepackage{amsmath,amssymb}
\usepackage{graphicx}% Include figure files
\usepackage{dcolumn}% Align table columns on decimal point
\usepackage{bm}% bold math
\usepackage{color}

\begin{document}

\preprint{APS/123-QED}

%\title{Intrinsic versus extrinsic cellular decision making}
\title{Quantifying cellular autonomy in multi-cue environments}

\author{Louis Gonz\'alez}
\affiliation{Department of Physics and Astronomy, University of Pittsburgh, Pittsburgh, PA 15260, USA}
\author{Hogyeong Gwak}
\affiliation{Department of Mechanical Science and Engineering, University of Illinois, Urbana-Champaign, Urbana, IL 61801, USA}
\author{Bumsoo Han}
\affiliation{Department of Mechanical Science and Engineering, University of Illinois, Urbana-Champaign, Urbana, IL 61801, USA}
\affiliation{Cancer Center at Illinois, University of Illinois Urbana-Champaign, Urbana, IL 61801, USA}
\affiliation{Chan Zuckerberg Biohub Chicago, Chicago, IL 60642, USA}
\author{Andrew Mugler}
\email{andrew.mugler@pitt.edu}
\affiliation{Department of Physics and Astronomy, University of Pittsburgh, Pittsburgh, PA 15260, USA}

\begin{abstract}
A cell routinely responds to one of many competing environmental cues. A fundamental question is whether the cell follows the cue prioritized by its internal signaling network or the cue that carries the most external information. We introduce a theoretical framework to answer this question. We derive information limits for four types of directional cues: external and self-generated chemical gradients, fluid flow, and contact inhibition of locomotion. When the cues compete as pairs, these limits predict which cue a cell should follow if its decision is based on environmental information alone. We compare these predicted decision boundaries with data from our and others' cell migration experiments, finding cases where the boundary is obeyed and cases where it is violated by orders of magnitude. Both outcomes are informative, and we find that they rationalize known properties, or predict putative properties, of cells' internal signaling networks. Our work introduces a physical framework to quantify the degree to which a cell acts like an autonomous agent, rather than a passive detector, favoring a cue even when it is less informative.
\end{abstract}

\maketitle

\section{Introduction}
Cells navigate their environment by interpreting multiple, often competing, signals—chemical gradients, mechanical forces, and interactions with other cells. A key question is how cells integrate these cues to make decisions. One approach emphasizes the cell’s signaling network, which determines how a cue is processed internally \cite{hartwell1999molecular, alon2019introduction, saha2022deduction}. Another approach considers the external information available in each cue, asking whether a cell's behavior approximates that of an optimal physical detector \cite{Berg1977, endres2008accuracy, Bouffanais2013, mugler2016limits, Fancher2020, wang2024limits}. When signals conflict, does a cell follow the cue prioritized by its signaling network, or the cue that carries the most information? Here, we make this question quantitative and test it experimentally.

We focus on directed cell migration, a fundamental process in development \cite{Ewald2008}, wound healing \cite{Phillipson2011}, and cancer metastasis \cite{Friedl2003}. Migrating cells respond to chemical gradients \cite{kim2013cooperative, moon2021signal, Moon2023}, including self-generated ones \cite{Shields2007}, as well as mechanical signals like shear \cite{decave2003shear, sadhukhan2023modelling} or pressure gradients \cite{Polacheck2011, Polacheck2014} induced by fluid flow. They also interact with other cells through contact inhibition of locomotion, where contact reverses migration \cite{mayor2010keeping, lin2015interplay}. In complex environments such as tumors, multiple cues are present simultaneously, making the problem of cue selection both fundamental and physiologically relevant.

\begin{figure}	
  \includegraphics[width=\columnwidth]{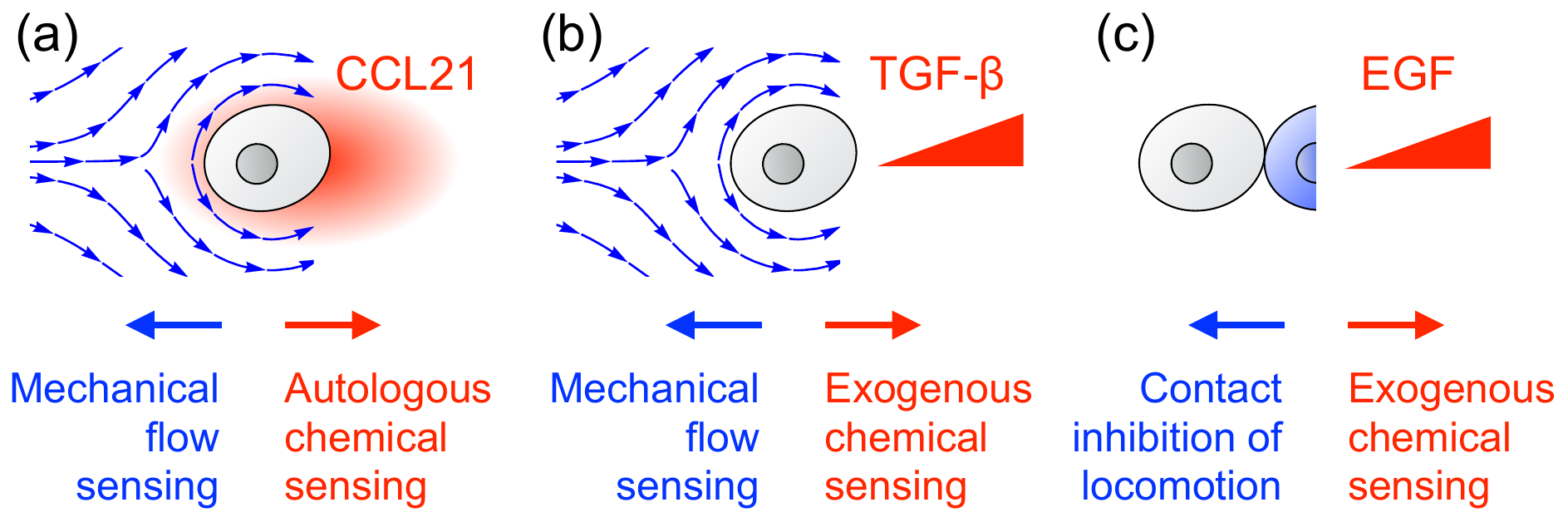}
  \caption{	\label{cartoon} The three experimental examples we investigate, where competing cues promote opposite migratory outcomes for a cell.}
\end{figure}

We investigate three experimental scenarios in which cells must respond to one of two competing cues (Fig.\ \ref{cartoon}): (a) an autologous (self-generated) chemical gradient vs.\ a flow-induced pressure gradient, (b) an exogenous chemical gradient vs.\ a flow-induced pressure gradient, and (c) an exogenous chemical gradient vs.\ contact inhibition of locomotion from another cell. We consider the information contained in each signal, deriving physical bounds on the precision with which each could be sensed. This approach leads to decision boundaries separating phases where the cell is predicted to follow one cue over the other if the cell optimally detects and equally weighs the cues in its decision. We then compare these predictions with data from experiments, both our own and previously published. Because all free parameters of the theory have been measured in these experiments, the theory makes falsifiable predictions that the data either do or do not obey.

We find that the decision boundary is obeyed in case (a), partially obeyed in case (b), and strongly violated in case (c). These findings have important implications for the cellular decision-making process in each case. In case (a), for example, the cell's decision is equivalent to that of a passive, comparative detector, whereas in case (c) the cell's decision reflects a strong preference for a one cue over another, even when it has the smaller information content. We map these outcomes to the known features of the cell's signaling network in each case. Where these features are known in detail, our findings provides rationales for their function. Where they are not known, our findings make predictions for their structure. Altogether, our results introduce and demonstrate a framework for quantifying cellular autonomy: the degree to which a cell favors a cue even when it is less informative.

\section{Results}

We begin by defining the detectable information in four types of environmental cues and deriving physical limits to the detection precision. In each case, we take precision to be the signal-to-noise ratio for a physical quantity that contains the directional information of the cue. Our theory is built on simple scaling arguments because we find that when decision boundaries are violated, they are violated by orders of magnitude. In all cases, we will find that our results agree with previous results derived more rigorously but in special cases. This validates our scaling approach, even as we go beyond these cases. Throughout this work we assume spherical cells with uniform, noninteracting sensors, and we consider relaxations of these assumptions in the Discussion.

\subsection{Autologous chemical sensing vs.\ mechanical flow sensing}
In the presence of a slow background fluid flow, multiple cancer cell types have been shown to secrete and detect a chemical called CCL21 whose distribution is biased by the flow \cite{Shields2007, munson2013interstitial}. This creates a downstream gradient that cells track, a process termed autologous chemotaxis \cite{Shields2007}. However, at high cell density, these cells instead migrate upstream due to mechanical sensing of the flow-induced pressure gradient across the cell body \cite{Polacheck2011}. The hypothesis is that high cell density corresponds to a large background concentration of the secreted chemical, saturating the chemical gradient and allowing pressure sensing to take over \cite{Polacheck2011, Vennettilli2022}. Despite important theoretical work on this problem \cite{Vennettilli2022, khair2021two, ben2024using}, this hypothesis has not been quantitatively tested.

\subsubsection{Autologous chemical sensing}
\begin{figure}
  \includegraphics[width=.9\columnwidth]{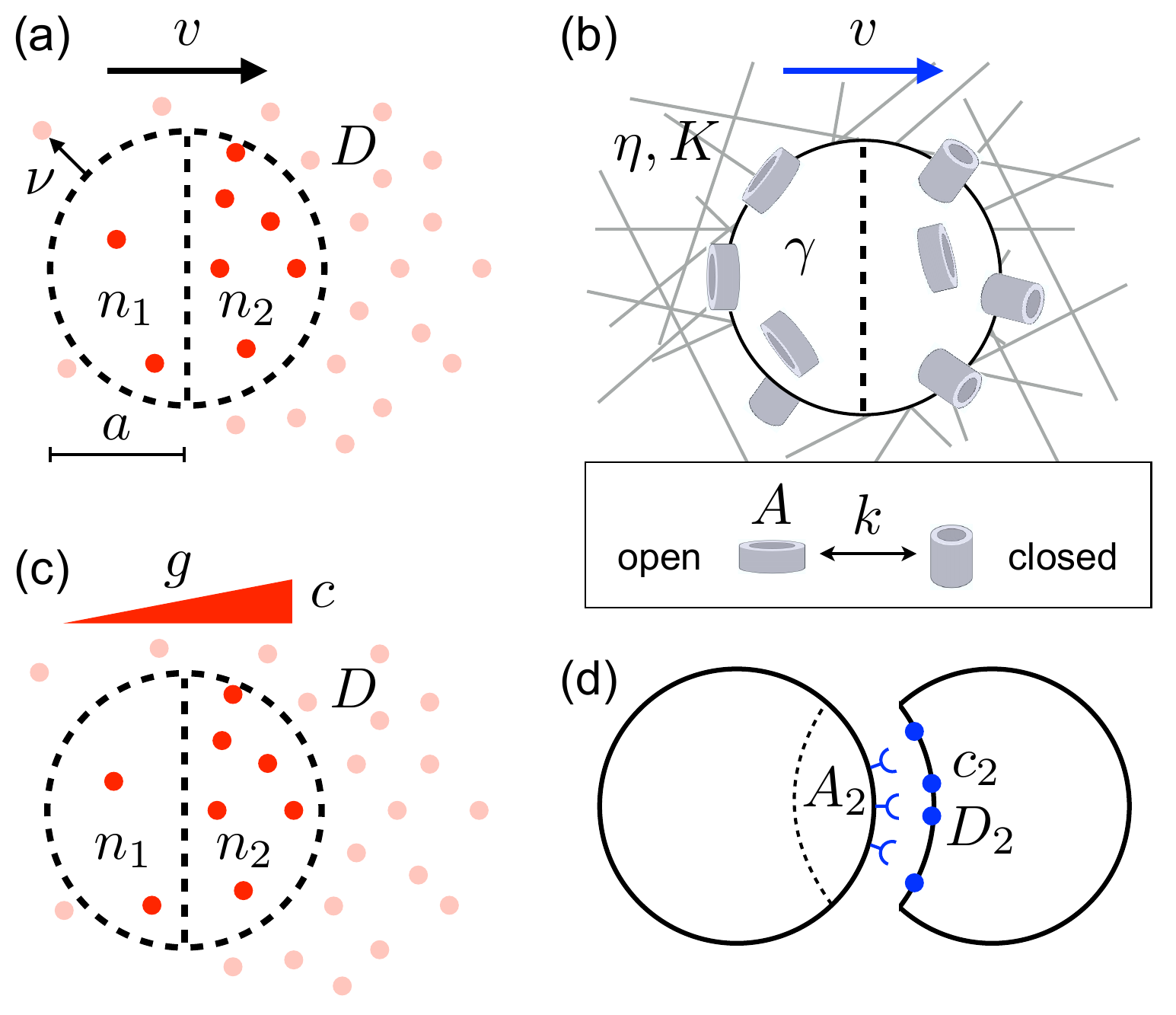}
  \caption{ \label{sensing} Schematics of (a) autologous chemical sensing, (b) mechanical flow sensing, (c) exogenous chemical sensing, and (d) contact inhibition of locomotion. See text and Table I for parameter descriptions.}
\end{figure}

We first derive the physical limit to the precision of chemical gradient sensing for this process. The directional information is contained in the difference $\Delta n = n_2-n_{1}$ in the number of molecules that could be detected by the two halves of the cell [Fig.\ \ref{sensing}(a)] \cite{mugler2016limits, Vennettilli2022, Gonzalez2023}. For slow flow, the mean of this difference scales as $\Delta\bar{n} \sim \epsilon\nu a^2/D$ \cite{Vennettilli2022}, where $a$ is the cell lengthscale, $\nu$ is the molecule secretion rate, $D$ is the molecular diffusion coefficient, and the flow speed $v$ enters through the Pecl\'et number $\epsilon = va/D \ll 1$ (see Table I for a list of all parameters in this work, as well as their experimentally measured values). This expression for $\Delta\bar{n}$ follows from taking, in each half of the cell, the ratio of the rate of molecule gain due to secretion, to the rate of loss due to diffusion and flow \cite{Vennettilli2022}.

\begin{table}[b]
\begin{center}
\begin{tabular}{|l|l|l|l|}
\hline
& {\bf Meaning} & {\bf Value} & {\bf Reference} \\
%\hline
\hline
$a$ & Cell lengthscale & 10 $\mu$m & \cite{Polacheck2011, Polacheck2014, Moon2023, lin2015interplay} \\
\hline
$\nu$ & Molecule secretion & 1 s$^{-1}$ & \cite{Shields2007}: Fig 3(a) main\\
 & & $6$ s$^{-1}$ & \cite{munson2013interstitial}: Fig 3(a) inset \\
\hline
$D$ & Molecule diffusion & 150 $\mu$m$^2$/s & \cite{Fleury2006, munson2013interstitial} \\
\hline
$L$ & Chamber length & 3 mm & \cite{Polacheck2011, haessler2012migration} \\
\hline
$K$ & Permeability & $0.1$ $\mu$m$^2$ & \cite{Polacheck2011}: Fig 3(a) \\
 & & $0.05$ $\mu$m$^2$ & \cite{Moon2023}: Fig 3(b) \\
\hline
$\Delta A$ & Channel area change & $6.5$ nm$^{2}$ & \cite{sukharev1999energetic} \\
\hline
$N$ & Channels per cell & $1000$ & \cite{morris1990} \\
\hline
$k$ & Channel switching rate & $1$ s$^{-1}$ & \cite{shapovalov2004gating} \\
\hline
$\rho$ & Cell density & Fig 3(a) & \cite{Polacheck2011, Polacheck2014, haessler2012migration, munson2013interstitial} \\
\hline
$v$ & Fluid flow speed & Fig 3(a, b) & \cite{Polacheck2011, Polacheck2014, haessler2012migration, munson2013interstitial, Moon2023}, \\
& & & this work \\
\hline
$c$ & Background conc. & Fig 3(b, c) & \cite{Moon2023, lin2015interplay}, this work \\
\hline
$g$ & Concentration gradient & Fig 3(b, c) & \cite{Moon2023, lin2015interplay}, this work \\
\hline
$c_2$ & Ephrin concentration & Fig 3(c) & \cite{xu2011epha2} \\
\hline
$D_2$ & Ephrin diffusion & Fig 3(c) & \cite{xu2011epha2} \\
\hline
$\tau$ & Integration time & Drops out & \\
\hline
$A_2$ & Cell-cell contact area & Drops out & \\
\hline
$\eta$ & Viscosity & $10^{-3}$ Pa$\cdot$s & Water \\
\hline
$\beta$ & Inverse temperature & $2.5$$\times$$10^{20}$/J & Room temp. \\
\hline
\end{tabular}
\caption{Parameters used in this work. See Appendices \ref{theory} and \ref{data} for details on how their values are measured.}
\label{params}
\end{center}
\end{table}

The variance in $\Delta n$ scales the same way as the variance in the total number $n$ of molecules detected by the cell \cite{mugler2016limits}. Because diffusion causes the number of molecules in any finite volume to follow a Poisson distribution, this variance equals its mean $\bar{n}$. In a time $\tau$, the variance can be reduced as $\sigma^2_{\Delta n} \sim \bar{n}/M$ by the cell integrating $M = D\tau/a^2$ independent measurements, where $M$ follows from dividing $\tau$ by the diffusive correlation time $a^2/D$ \cite{Berg1977}. The mean scales as $\bar{n} \sim \nu a^2/D + ca^3$, where the first term corresponds to molecules secreted by the cell, and the second term corresponds to a background concentration $c$ of molecules secreted by other cells \cite{Vennettilli2022}. The background concentration is related to the cell density $\rho$ by balancing molecule secretion with molecule loss from flow, giving $c = (\nu L/v)\rho$, where $L$ is the length of the experimental environment in the flow direction \cite{Vennettilli2022}.

Putting these results together, we arrive at the precision of autologous chemical sensing,
\begin{equation}
\label{Pchem1}
{\cal P}_{\rm chem}^{\rm auto} = \frac{\Delta \bar{n}}{\sigma_{\Delta n}} \sim \epsilon\sqrt{\frac{\epsilon\nu\tau}{\epsilon + \rho a^2L}},
\quad {\rm with} \quad \epsilon = \frac{va}{D}.
\end{equation}
This expression demonstrates that precision increases with faster flow, via $\epsilon$, but decreases with higher cell density $\rho$. Although it is derived using simple scaling arguments, these scalings are validated in special cases by more rigorous results derived previously using a reaction diffusion treatment that accounts explicitly for the fluid flow field \cite{Fancher2020, Vennettilli2022}. Specifically, $\Delta \bar{n}/\sigma_{\Delta n} = \epsilon\sqrt{\nu\tau}$ with $\rho=0$ agrees with Eq.\ 15 in \cite{Fancher2020}, and $\Delta\bar{n}/\bar{n} = \epsilon^2/(\epsilon+\rho a^2L)$ agrees with Eq.\ 8 in \cite{Vennettilli2022}, up to factors of order unity.

\subsubsection{Mechanical flow sensing}
We now seek to compare this result with the physical limit to the precision of mechanical sensing of the flow. Evidence suggest that cells detect flow using mechanosensitive ion channels. For example, amoeba cells move in response to flow in a calcium-dependent manner \cite{fache2005calcium}, and their speed decreases when mechanosensitive calcium channels are blocked \cite{lombardi2008mechano}. Similarly, the movement of keratocytes (eye cells) has been shown to depend on mechanosensitive calcium channels \cite{lee1999regulation}. Mechanosensitive ion channels are actuated by changes in membrane tension \cite{luchtefeld2024dissecting}. Experiments have demonstrated that these tension changes remain locally confined because relaxation of the membrane is impeded by transmembrane proteins bound to the cytoskeleton \cite{shi2018cell}.

Previous work derived the precision of flow detection when ion channels respond to changes in membrane tension caused by a flow-induced shear \cite{Bouffanais2013}. However, when cells are embedded in a dense extracellular matrix, as in the experiments considered here \cite{Polacheck2011, Polacheck2014, haessler2012migration, munson2013interstitial}, the flow is low-permeability, meaning that the average pore size of the matrix is much smaller than the cell. At low permeability, the force on the cell due to the flow is dominated by pressure, not shear \cite{Polacheck2014}. Therefore we adapt the previous treatment to pressure sensing. Again, we focus on simple scaling arguments, which to our knowledge are absent from the literature for this process.

The flow induces a pressure difference $\Delta P$ across the cell. At low permeability $K \ll a^2$, the pressure difference scales with the flow velocity $v$ as $\Delta P \sim v\eta a/K$ \cite{Polacheck2014}, where $\eta$ is the fluid viscosity. The pressure difference causes a membrane tension difference via the Young-Laplace equation, $\Delta\gamma \sim a\Delta P$. As in the previous work \cite{Bouffanais2013}, we suppose that the cell detects this tension difference using $N$ ion channels distributed across the membrane [Fig.\ \ref{sensing}(b)]. Each channel switches between an open and a closed conformation. The probability $p$ that the channel is open obeys that for a two-state system in a thermal bath, $p = (1+e^{-q})^{-1}$, where $q = \beta\Delta E$ is the energy difference $\Delta E$ between the closed and open states, scaled by the inverse temperature $\beta$. The energy difference is given by $\Delta E = \Delta\gamma\Delta A$, where $\Delta A$ is the difference in channel area between the closed and open state \cite{Bouffanais2013, ursell2008role}. Thus, an increase in pressure on the upstream side of the cell causes an increase in membrane tension, which increases the probability that a channel there is open.

Just as the directional information of the chemical cue was contained in the difference in the detected molecule numbers from the two halves of the cell, here the directional information of the mechanical cue is contained in the difference in the dimensionless tension readouts $q$ from channels on the two halves of the cell. (The result is equivalent if we instead take the readout to be the difference in open probabilities $p$; see Appendix \ref{alt}.) On the upstream half the $N/2$ channels will read out a tension increase, whereas on the downstream half the $N/2$ channels will read out a tension decrease. Thus, the mean difference will scale as $\bar{Q}\sim (N/2)q - (N/2)(-q) = Nq$. The variance, assuming the channels are sufficiently independent \cite{Bouffanais2013}, will scale as $\sigma_Q^2 \sim N\sigma_q^2/M$, where here $M=(k_++k_-)\tau$ because the correlation time is the inverse of the sum of the channel switching rates $k_+$ and $k_-$. The variance in $q$ propagates in quadrature from $p = (1+e^{-q})^{-1}$ as $\sigma_q^2 = 1/\sigma_p^2$ (Appendix \ref{alt}), where $\sigma_p^2 = \bar{p}(1-\bar{p})$ for a two-state switch. The minimum noise occurs when the channels have $\bar{p} = 1/2$ (and thus $k_+=k_-\equiv k$) in the absence of flow, a condition called critical pre-stress \cite{Bouffanais2013}.

Putting these results together, we arrive at the precision of mechanical flow sensing,
\begin{equation}
\label{Pmech}
{\cal P}_{\rm mech} = \frac{\bar{Q}}{\sigma_Q} \sim q\sqrt{Nk\tau},
\quad {\rm with} \quad q = \frac{\beta\eta va^2\Delta A}{K}.
\end{equation}
This expression demonstrates that precision increases with faster flow, via $q$. Again, although it is derived using simple scaling arguments, it agrees with the previous work \cite{Bouffanais2013} in the case that tension change is induced by shear instead of pressure. Specifically, Eq.\ \ref{Pmech} agrees with Eqs.\ 41 and 42 in \cite{Bouffanais2013} up to factors of order unity when (i) the tension change $\Delta\gamma = a\eta {\cal G}$ arises from the shear rate ${\cal G}$ rather than the pressure difference, (ii) critical pre-stress is assumed ($\Delta h-\gamma_0\Delta A = 0$ there), and (iii) the membrane stretch energy (much smaller than the tension energy) is ignored. Note that $k$ here then corresponds to $k_0 e^{-\beta\gamma_0\Delta A/2}$ there, which given the parameter choices there evaluates to about once per second, consistent with the measured value we use here (Table I).

\subsubsection{Comparing with data}
\begin{figure*}
  \includegraphics[width=\textwidth]{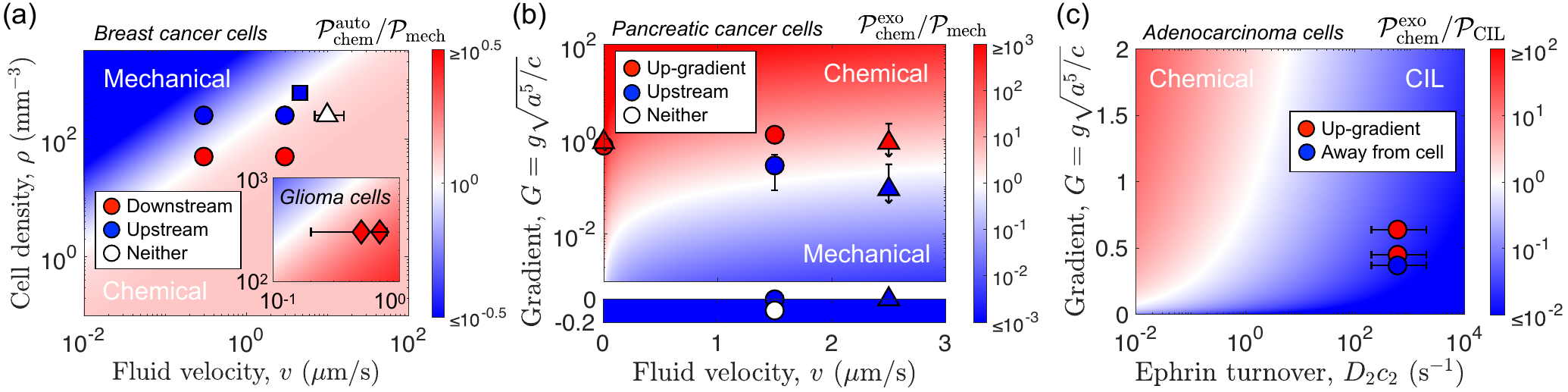}
  \caption{ \label{phase} Predicted decision boundaries (color maps) and observed cell behavior (data points) for the three cases in Fig.\ \ref{cartoon}. Agreement indicates that decisions follow the higher-information cue, whereas violation indicates that decisions are autonomous---following a cue even when it is lower-information. Data are from \cite{Polacheck2011} (a, circles), \cite{Polacheck2014} (a, square), \cite{haessler2012migration} (a, triangle), \cite{munson2013interstitial} (a, diamonds), \cite{Moon2023} (b, circles), this work (b, triangles), and \cite{lin2015interplay} (c). In b, down arrows on triangles indicate that error window drops below zero. See Appendices \ref{theory} and \ref{data} for analysis details and \ref{exp} for experimental methods. See \cite{code} for code and data.}
\end{figure*}

All parameters in Eqs.\ \ref{Pchem1} and \ref{Pmech} have been measured experimentally (Table I). Therefore, we can evaluate the ratio ${\cal P}_{\rm chem}^{\rm auto}/{\cal P}_{\rm mech}$ with no free parameters. Figure \ref{phase}(a) shows this ratio as a function of flow speed and cell density (color map) for two cell types with different secretion rates $\nu$ (main panel and inset, respectively). We see that, as hypothesized, the precision of mechanical sensing overtakes that of chemical sensing as the cell density increases.

Overlaid as data points in Fig.\ \ref{phase}(a) are the measured flow speeds and cell densities of cells that are observed to travel upstream, downstream, or neither \cite{Polacheck2011, Polacheck2014} (see Appendix \ref{data} for details on all data points in Fig.\ \ref{phase}). We see upstream migration where mechanical sensing is predicted to dominate strongly (upper left point), downstream migration where chemical sensing is predicted to dominate strongly (inset), and a mix of responses near the boundary or where chemical sensing dominates only weakly (light red region). This finding suggests that these cells decide between autologous chemical sensing and mechanical flow sensing based on the information contained in the two environmental signals.

\subsection{Exogenous chemical sensing vs.\ mechanical flow sensing}
We previously investigated a cancer cell line exposed to both a fluid flow and a gradient of an exogenous chemical attractant called TGF-$\beta$ \cite{Moon2023}. We found that in locations where the gradient was downstream and steep, cells migrated downstream. However, in locations where the gradient was downstream but shallow, cells migrated upstream. Here we ask whether these observations are consistent with the precision crossover between exogenous chemical sensing and mechanical flow sensing.

\subsubsection{Exogenous chemical sensing}
Our derivation for exogenous chemical sensing follows that above for autologous chemical sensing, but now the average number $\bar{n}$ of detected molecules and the difference $\Delta\bar{n}$ between the two halves of the cell are set by the local background concentration $c$ and gradient $g$ of an exogenous chemical signal [Fig.\ \ref{sensing}(c)]. Specifically, we have $\bar{n} \sim ca^3$ and $\Delta\bar{n} \sim (ga)a^3$, where $ga$ is the approximate concentration difference across the cell \cite{mugler2016limits}. The other elements follow from above: $\sigma^2_{\Delta n} \sim \bar{n}/M$, with $M = D\tau/a^2$. Thus we arrive at the precision of exogenous chemical sensing,
\begin{equation}
\label{Pchem2}
{\cal P}_{\rm chem}^{\rm exo} = \frac{\Delta \bar{n}}{\sigma_{\Delta n}} \sim G\sqrt{\frac{D\tau}{a^2}},
\quad {\rm with} \quad G = g\sqrt{\frac{a^5}{c}}.
\end{equation}
The dimensionless parameter $G$ captures the fact that precision increases with the gradient but decreases with the background concentration: it is harder to measure a molecule number difference against larger background fluctuations \cite{mugler2016limits}. Once again, although this expression is derived using simple scaling arguments, it agrees with a more rigorous result derived previously using a reaction diffusion treatment that accounts for internal signaling within the detector. Specifically, Eq.\ \ref{Pchem2} agrees with Eq.\ 11 in \cite{mugler2016limits} up to factors of order unity in the limit of a single cell ($n_0 = 1$ there).

\subsubsection{Comparing with data}
Again, because all parameters in Eqs.\ \ref{Pmech} and \ref{Pchem2} have been measured experimentally (Table I), we can evaluate the ratio ${\cal P}_{\rm chem}^{\rm exo}/{\cal P}_{\rm mech}$ with no free parameters. Figure \ref{phase}(b) shows this ratio as a function of flow speed and dimensionless gradient strength $G$ (color map). We see that chemical sensing is more precise for larger gradients, while mechanical sensing is more precise for larger flow speeds. Overlaid as data points (circles) are the measured flow speeds and $G$ values from our previous experiments \cite{Moon2023} of cells that are observed to travel upstream, downstream, or neither. Although most points agree with the predicted decision boundary within error, the data are sparse. Therefore, we performed new experiments, focusing on the higher flow speed of $v = 2.5$ $\mu$m/s (see experimental methods in Appendix \ref{exp}).

\begin{figure*}
  \includegraphics[width=.75\textwidth]{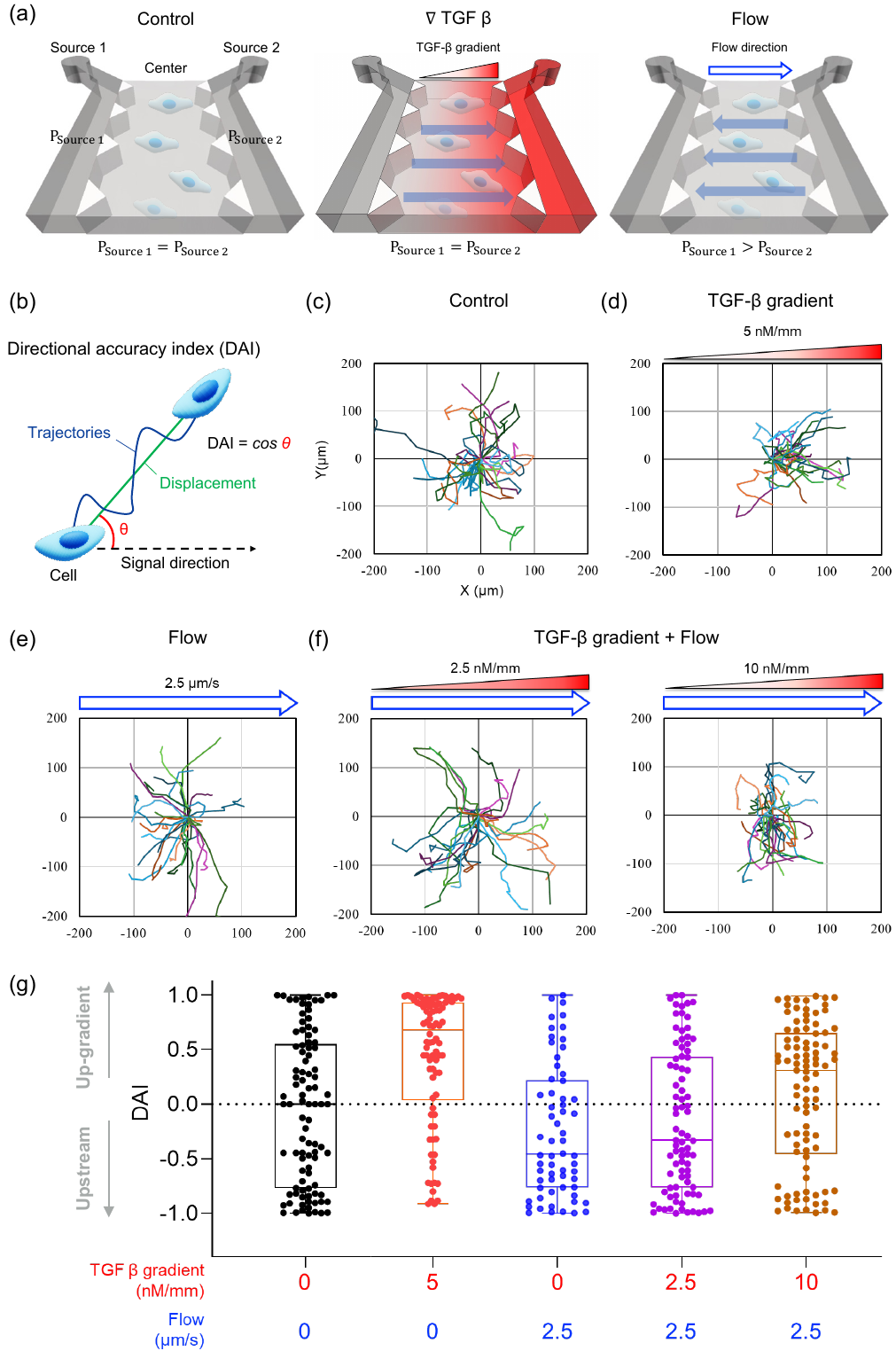}
  \caption{	\label{newdata} Experiments on cell migration with chemical and fluidic cues for Figure \ref{phase}(b) (triangles). (a) Schematic diagram of microfluidic platform. (b) Directional accuracy index. (c-f) Cell migration trajectories in five experimental conditions. (g) DAI distributions. Box: quartiles with median line. Dots: DAI values from cell trajectories ($N > 60$). Statistical comparisons were performed using the one-sample Wilcoxon ($p_{\rm W}$) test and the nonparametric Mann-Whitney test ($p_{\rm MW}$). Both the TGF-$\beta$-only condition (red, $p_{\rm W} < 0.0001$, $p_{\rm MW} < 0.0001$) and the flow-only condition (blue, $p_{\rm W} < 0.01$, $p_{\rm MW} = 0.0862$) are significantly different from the control (black), and the small- and large-gradient conditions with flow (purple, brown) are significantly different from each other ($p_{\rm W} = 0.0247$, $p_{\rm MW} = 0.0145$).}
\end{figure*}

Specifically, we exposed murine pancreatic cancer cells (KIC cells) to two cues, a TGF-$\beta$ gradient and fluid flow, in our microfluidic platform [Fig.\ \ref{newdata}(a)]. Cell migration was evaluated using the directional accuracy index (DAI), the cosine of the angle between a cell's displacement and the cue direction [Fig.\ \ref{newdata}(b)]. Five experimental conditions were investigated: no growth factor and no flow (control), a 5 nM/mm TGF-$\beta$ gradient only, a 2.5 $\mu$m/s flow only, a small-gradient TGF-$\beta$ profile from $2.5$ to $5$ nM with a 2.5 $\mu$m/s flow, and a large-gradient TGF-$\beta$ profile from $0$ to $10$ nM with a 2.5 $\mu$m/s flow [Fig.\ \ref{newdata}(c-f)]. As seen in Fig.\ \ref{newdata}(g), whereas the median DAI is zero for the control (black), cells migrated up-gradient (DAI $>0$) in the gradient-only condition and upstream (DAI $<0$) in the flow-only condition. When the two cues competed, cells migrated upstream when the gradient was small (purple) and up-gradient when the gradient was large (brown).

The results in Fig.\ \ref{newdata}(g) are summarized by the triangles in Figure \ref{phase}(b), and we see that they show further agreement with the decision boundary. Together with the previous data (circles), Figure \ref{phase}(b) suggests that these cells decide between exogenous chemical sensing and mechanical flow sensing based on the information contained in the two environmental signals.

The notable exception to the agreement in Fig.\ \ref{phase}(b) is the white circle deep within the blue regime: these cells are predicted to migrate upstream but are observed, on average, to migrate neither upstream nor downstream.
The difference between this point and the others in Fig.\ \ref{phase}(b) is that it corresponds to a very high background concentration \cite{Moon2023}. In fact, here the gradient even points slightly upstream ($g<0$), meaning that, for this point only, both cues promote upstream migration with no conflict. Yet, the high background concentration evidently prevents cells from responding to either cue. This is in marked contrast to the previous case [Fig.\ \ref{phase}(a)]. There, the background concentration was also high (at high cell density), but there, the cells responded to one of the cues (mechanical flow). This contrast has important implications for the intrinsic signaling networks in these two cases, which we expand upon in the Discussion.

\subsection{Exogenous chemical sensing vs.\ contact inhibition of locomotion}
Previous work investigated a cancer cell line exposed to a gradient of an exogenous chemical attractant called EGF and placed in narrow channels \cite{lin2015interplay}. This setup caused many encounters where up-gradient migration conflicted with repulsion from another cell due to contact inhibition of locomotion (CIL). At low background EGF concentration, cells migrated up-gradient, whereas at high background concentration, cells migrated away from the other cell. Here we ask whether these observations are consistent with the precision crossover between exogenous chemical sensing and CIL.

\subsubsection{Contact inhibition of locomotion}
Recent work investigated the physical aspects of CIL accuracy \cite{wang2024limits} but focused on the effects of contact size and signal interference rather than a general precision limit. Here, as above, we derive the physical limit from simple scaling arguments. The mechanism of CIL is that each cell's membrane has molecules called ephrin, as well as receptors to detect that molecule on another cell \cite{lin2015interplay, wang2024limits} [Fig.\ \ref{sensing}(d)]. Calling the contact area $A_2$ and the two-dimensional concentration of ephrin $c_2$, the mean number of ephrin moelcules scales as $\bar{n} \sim c_2A_2$. Calling the diffusion coefficient of ephrin $D_2$, the variance in this number, as above, scales as $\sigma_n^2 \sim \bar{n}/M$ with $M = D_2\tau/A_2$. Thus we arrive at the precision of CIL,
\begin{equation}
\label{PCIL}
{\cal P}_{\rm CIL} = \frac{\bar{n}}{\sigma_n} \sim \sqrt{D_2c_2\tau}.
\end{equation}
We note that the contact area $A_2$ drops out, which is consistent with the finding in the previous work that accuracy is largely insensitive to contact size \cite{wang2024limits}.

\subsubsection{Comparing with data}
Figure \ref{phase}(c) shows the ratio ${\cal P}_{\rm chem}^{\rm exo}/{\cal P}_{\rm CIL}$ as a function of ephrin turnover $D_2c_2$ and dimensionless gradient strength $G$ (color map), again with all remaining parameters measured experimentally (Table I). We see that chemical sensing is more precise for larger gradients, while CIL is more precise for higher ephrin turnover. Overlaid as data points are the measured $G$ values from the narrow channel experiments \cite{lin2015interplay} and the values of $D_2c_2$ from experiments that measured the diffusion and concentration of ephrin \cite{xu2011epha2}. We see that the measured decision boundary is three orders of magnitude away from the predicted boundary. This finding suggests that these cells do not decide between exogenous chemical sensing and CIL based on signal information alone. Specifically, it suggests that cells' internal signaling networks weigh information about the EGF gradient over information from CIL.

\section{Discussion}
We have developed a framework for quantifying the degree to which a cell's decision between two environmental cues is consistent with the information content of those cues. The framework produces decision boundaries that predict how cells should respond if they weigh signal information without prejudice. Violation of these decision boundaries indicates that cells do have a prejudice---favoring a cue even when it provides less information---a concept we term cellular autonomy. We have applied this framework in three separate experimental examples with no free parameters. We have found that the results vary widely: depending on the cue types, the respective decision boundaries are obeyed, partially obeyed, or strongly violated (Fig.\ \ref{phase}).

\begin{figure}	
  \includegraphics[width=\columnwidth]{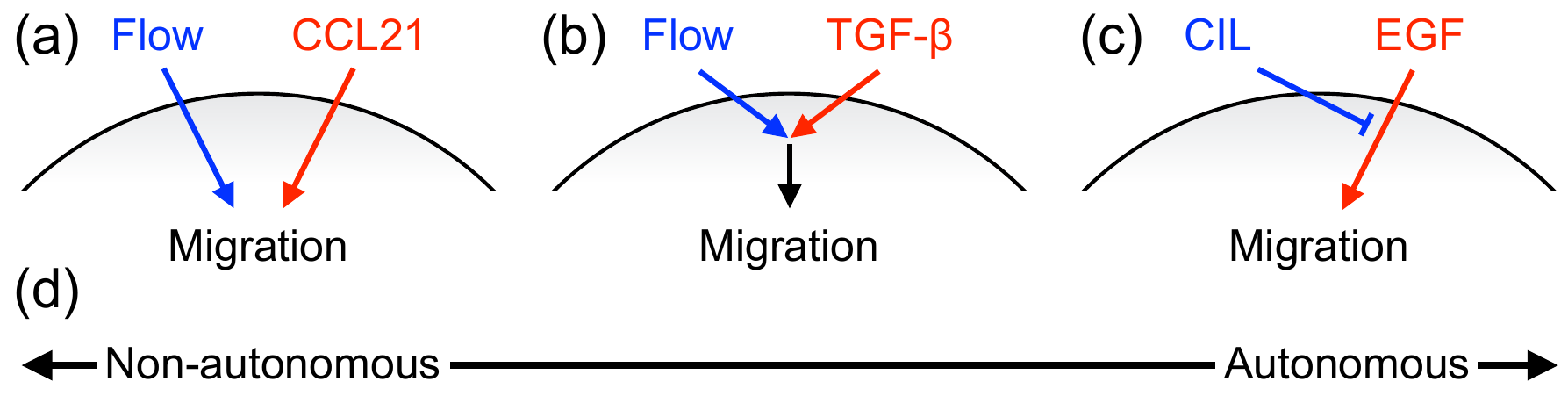}
  \caption{	\label{networks} Pathway structure (a) that we predict or (b, c) that is previously known from experiments. (d) The three cases define a spectrum from non-autonomous to autonomous decision-making.}
\end{figure}

What are the implications of Fig.\ \ref{phase} for the cell's signaling network in each case? The strong violation of the decision boundary in Fig.\ \ref{phase}(c) suggests that the EGF pathway and the CIL pathway interact upstream of the migration response, rather than transmitting the signal information to the migration response independently. In fact, in the previous experimental work on this case \cite{lin2015interplay}, it was demonstrated that the ephrin receptor represses the activity of the EGF receptor directly, far upstream of the cascade that connects this activity to the migration response [Fig.\ \ref{networks}(c)]. The cell could easily tune the strength of this repression, with weaker repression suppressing information from the CIL signal and shifting the decision boundary further into the naively CIL-dominated regime, as in Fig.\ \ref{phase}(c).

Similarly, the violation of the decision boundary at high background concentration in Fig.\ \ref{phase}(b) (white circle) but not Fig.\ \ref{phase}(a) suggests that the TGF-$\beta$ pathway and the flow sensing pathway converge upstream of the migration response. In fact, in our previous experimental work, we found that the results in Fig.\ \ref{phase}(b), as well as many other results including cases of commensurate (rather than conflicting) cues, are explained by a model in which the two pathways converge on a common network component whose activity can saturate \cite{moon2021signal, saha2022deduction, Moon2023} [Fig.\ \ref{networks}(b)]. Conversely, while to our knowledge the interaction between the CCL21 pathway and the flow sensing pathway is less characterized, our results in Fig.\ \ref{phase}(a) suggest that these two pathways transmit the signal information to the migration response faithfully and independently [Fig.\ \ref{networks}(a)]. Taken together, our results imply a spectrum of non-autonomous to autonomous decision-making for these three cases [Fig.\ \ref{networks}(d)], with cases (a), (b), and (c) progressively more autonomous, respectively.

Throughout this work we have assumed uniform, non-interacting sensors on spherical cells. What are the potential implications of relaxing these assumptions? With regard to sensor distribution, it is likely that both the receptors and the ion channels are well distributed across the cell surface, even during sensing. For example, the CCR7 receptors of autologous chemotaxis, as well as their bound ligands, are seen in microscopy images to be uniformly distributed (see Figs.\ 2A and 3A of \cite{Shields2007}). Mechanosensitive ion channels are observed to be uniformly distributed in many cell types, with an average spacing (1 $\mu$m) that is much larger than their size (see \cite{morris1990} and references therein). Even if receptors or channels are well distributed but ``patchy,'' previous work has demonstrated that this feature has little effect on sensory precision \cite{Fancher2020}.

With regard to sensor interactions, nonlinear interactions could amplify or otherwise change the sensory response \cite{adler2018fold}. Importantly, however, such effects would amount to a local modification of the input signal. According to the data processing inequality \cite{cover2005elements}, such a modification could not increase the amount of sensory information beyond that present in the original signal. Therefore, the precisions that we calculate should be seen as upper bounds to the amount of external information sensed. Because we are interested in external information contained in competing sensory signals, comparing their upper bounds is a fair way to construct a predictive theory.

Lastly, with regard to cell shape, the most prominent shape change observed during 3D migration is cell elongation in the migration direction \cite{Polacheck2011}. Previous work has found that ellipsoidal elongation can increase sensory precision by a modest amount, about $30\%$ for a cell that doubles its length \cite{Fancher2020}. Elongating in the signal direction would be expected to increase both the molecule number difference in chemical gradient sensing and the pressure difference in flow sensing, such that any effect on the precision ratio may be suppressed.

In this work, by focusing on the physics of sensory precision, we have made the question of how cells respond to competing cues quantitative and falsifiable. Our approach can be extended straightforwardly to other cue types, to alternative cue pairings, and to competitions among more than two cues, or more than two responses. Our framework predicts the signals that cells transmit faithfully, and the signals that cells may suppress or ignore in favor of others. It facilitates the interpretation of signaling networks and may guide the discovery of pathways that are still poorly characterized.

\section{Acknowledgments}
We thank Hye-ran Moon and Andre Levchenko for helpful discussions. All code and data (Figs.\ \ref{phase} and \ref{newdata}) are freely available \cite{code}. L.\ G.\ and A.\ M.\ were supported by National Science Foundation Grants No.\ PHY-2118561 and No.\ MCB-2118037. H.\ G.\ and B.\ H.\ were supported by National Institutes of Health Grants No.\ R01 CA254110 and No.\ U01 CA274304. B.\ H.\ was supported by National Institutes of Health Grant No.\ R33 HL159948.

\appendix

\section{Parameter values for theory}
\label{theory}

All parameters used for the theory color maps in Fig.\ \ref{phase} are listed in Table \ref{params}. Their values are taken from published experimental studies, as follows:
\begin{itemize}
\item The experiments use MDA-MB-231 breast cancer cells \cite{Polacheck2011, Polacheck2014, haessler2012migration}, RT2 glioma cells \cite{munson2013interstitial}, KIC pancreatic cancer cells \cite{Moon2023}, and MTLn3-B1 adenocarcinoma cells \cite{lin2015interplay}, all of which have lengthscale on the order of $a=10$ $\mu$m (see microscopy images therein).
\item One thousand MDA-MB-231 cells secrete 1.5 pg of CCL21 or CCL19 molecules (molecular weight around 10 kDa) in a 24 hour period (see Fig.\ 3F in \cite{Shields2007}), corresponding to a rate of $\nu = 1$ s$^{-1}$ per cell. One thousand RT2 cells secrete 5.75 pg of CXCL12 molecules (molecular weight around 10 kDa) in a 16 hour period (see Fig.\ 3B in \cite{munson2013interstitial}), corresponding to a rate of $\nu = 6$ s$^{-1}$ per cell.
\item The diffusion coefficients of CCL21 and CXCL12 in extracellular matrix were estimated to be around $D=150$ $\mu$m$^2$/s \cite{Fleury2006, munson2013interstitial}. The molecular weights of TGF-$\beta$ and EGF are similar (around 10 kDa), and therefore we use the same value of $D$.
\item The chamber length in the flow direction for the autologous chemotaxis experiments was approximately $L = 3$ mm (see Fig.\ 1A in \cite{Polacheck2011} and Fig.\ 1E in \cite{haessler2012migration}).
\item The permeability of 2 mg/mL type I collagen matrix was measured to be $K = 0.1$ $\mu$m$^2$ in \cite{Polacheck2011}. The same density of type I collagen matrix was used in \cite{Moon2023}, and the permeability was estimated at a similar value of $K = 0.05$ $\mu$m$^2$.
\item The conformational area change of mechanosensitive ion channel MscL was measured at $\Delta A = 6.5$ nm$^2$ \cite{sukharev1999energetic}.
\item Experiments find that mechanosensitive ion channels occur at a density on the order of $1$ $\mu$m$^{-2}$ that is notably uniform across the cell surface and ubiquitous across cell types (see \cite{morris1990} and references therein). Given a surface area of $4\pi a^2$ with $a=10$ $\mu$m, we have on the order of $N = 1000$ channels.
\item The residency times of mechanosensitive ion channels MscL and MscS were measured at approximately $1$ s (see the top panels of Figs.\ 2 and 4A in \cite{shapovalov2004gating}), giving a switching rate on the order of $k = 1$ s$^{-1}$.
\end{itemize}

\section{Flow sensing alternate derivation}
\label{alt}
In the main text we quantify the directional information in mechanical flow sensing by the difference in the dimensionless tension readouts $q$ from channels on the two halves of the cell. Here we show that the sensing precision is equivalent if we instead take the readout to be the difference in the probabilities $p$ for channels to be open on the two halves of the cell.

The average difference in open probabilities between the two halves of the cell will scale as
\begin{equation}
\bar{\Pi} = (N/2)(\bar{p}+\Delta p) - (N/2)(\bar{p}-\Delta p)
= N\Delta p,
\end{equation}
where $N$ is the number of ion channels, and $\Delta p$ is the change in probability due to the tension change. The variance will scale as
\begin{equation}
\sigma_\Pi^2 \sim \frac{N\sigma_p^2}{M},
\end{equation}
where $\sigma_p^2 = p(1-p)$ is the variance for an individual channel, and $M$ is the number of independent measurements a channel makes. Together, this gives a precision
\begin{equation}
\label{Pi_prec}
\frac{\bar{\Pi}}{\sigma_\Pi} \sim \frac{\Delta p \sqrt{NM}}{\sigma_p}.
\end{equation}
Using $p = (1-e^{-q})^{-1}$, we find
\begin{equation}
\label{delp}
\Delta p = \left.\frac{dp}{dq}\right|_{\bar{q}} q
	= \left[\frac{e^{-\bar{q}}}{(1+e^{-\bar{q}})^2}\right]q
	= \bar{p}(1-\bar{p})q
	= \sigma_p^2q.
\end{equation}
Similarly,
\begin{equation}
\label{varp}
\sigma_p^2 = \left(\left.\frac{dp}{dq}\right|_{\bar{q}}\right)^2\sigma_q^2
	= \sigma_p^4\sigma_q^2,
\end{equation}
Solving Eq.\ \ref{varp} for $\sigma_p$ and inserting into Eq.\ \ref{delp}, we obtain
\begin{equation}
\sigma_p = \frac{1}{\sigma_q}, \qquad
\Delta p = \frac{q}{\sigma_q^2}.
\end{equation}
Thus, Eq.\ \ref{Pi_prec} becomes
\begin{equation}
\frac{\bar{\Pi}}{\sigma_\Pi} \sim \frac{(q/\sigma_q^2) \sqrt{NM}}{1/\sigma_q}
	= \frac{q\sqrt{NM}}{\sigma_q}.
\end{equation}
This result is equivalent to
\begin{equation}
\frac{\bar{Q}}{\sigma_Q} = \frac{Nq}{\sqrt{N\sigma_q^2/M}}
	= \frac{q\sqrt{NM}}{\sigma_q}
\end{equation}
from the main text.

\section{Data points}
\label{data}

All data points plotted in Fig.\ \ref{phase} are taken from published experimental studies or our own experiments here, as summarized in Table \ref{params}. See \cite{code} for the code and data.

Specifically, for Fig.\ \ref{phase}(a):
\begin{itemize}
\item The cell seeding densities used in \cite{Polacheck2011} are $\rho = 50$ mm$^{-3}$ and $\rho = 250$ mm$^{-3}$. The cell seeding density used in \cite{Polacheck2014} is $\rho = 600$ mm$^{-3}$. The cell seeding density used in \cite{haessler2012migration} is $\rho = 250$ mm$^{-3}$. The cell seeding density used in \cite{munson2013interstitial} is $\rho = 300$ mm$^{-3}$.
\item The flow speeds applied in \cite{Polacheck2011} are $v = 0.3$ $\mu$m/s and $v= 3$ $\mu$m/s. The flow speed applied in \cite{Polacheck2014} is $v = 4.6$ $\mu$m/s. The flow speed applied in \cite{haessler2012migration} is $v = 7$$-$$16$ $\mu$m/s (see Fig.\ 2D therein). The flow speed applied in \cite{munson2013interstitial} is $v = 0.7$ $\mu$m/s or, in their pillar flow chamber, $v = 0.2$$-$$0.8$ $\mu$m/s.
\item In \cite{Polacheck2011}, at $\rho = 50$ mm$^{-3}$, cells migrated downstream on average for both flow speeds (see gray squares in Fig.\ 3B therein). In \cite{Polacheck2011}, at $\rho = 250$ mm$^{-3}$, cells migrated upstream on average for both flow speeds (see gray circles in Fig.\ 3B therein). In \cite{Polacheck2014}, cells migrated upstream on average (see lower left plot in Fig.\ 5C therein). In \cite{haessler2012migration}, distinct subpopulations of cells migrated downstream and upstream (see Fig.\ 5E therein), such that the total population did neither on average (see Fig.\ 3D therein, rightmost plot). In \cite{munson2013interstitial}, cells polarized downstream in their radial flow chamber (see Fig.\ 4B therein) and migrated downstream in their pillar flow chamber (see Fig.\ 5F-H therein).
\end{itemize}
For Fig.\ \ref{phase}(b), circles:
\begin{itemize}
\item The flow speed applied in \cite{Moon2023} is $v = 1.5$ $\mu$m/s.
\item The gradient $g$ and average background concentration $c$ are computed between all pairs of data points in \cite{Moon2023} with (i) no flow (Fig.\ 2D in \cite{Moon2023}), (ii) a large downstream gradient (right region of Fig.\ 2F in \cite{Moon2023}), and (iii) a small upstream gradient (right region of Fig.\ 2E in \cite{Moon2023}). The values of $G = g\sqrt{a^5/c}$ are calculated with $a=10$ $\mu$m and averaged. The standard deviations in $G$ are smaller than the data points in Fig.\ \ref{phase}(b).
\item For the case of (iv) a small downstream gradient (left region of Fig.\ 2F in \cite{Moon2023}), because the concentration values are so small, we use the same procedure as above, but for the raw concentration data rather than the trial- and window-averaged data in Fig.\ 2F of \cite{Moon2023}, for better accuracy. The error bar on this point in Fig.\ 3(b) is the standard deviation across six experimental trials of the average $G$ value in this region.
\item For the above cases, cells migrated, on average, (i) up-gradient (Fig.\ 3C in \cite{Moon2023}, second box plot), (ii) up-gradient (Fig.\ 3C in \cite{Moon2023}, sixth box plot), (iii) nowhere (Fig.\ 3C in \cite{Moon2023}, fifth box plot; compare with control), and (iv) upstream (Fig.\ 3C in \cite{Moon2023}, seventh box plot). In the case of flow with no gradient, cells migrated upstream on average (Fig.\ 3C in \cite{Moon2023}, third box plot).
\end{itemize}
For Fig.\ \ref{phase}(b), triangles (details in Appendix \ref{exp}):
\begin{itemize}
\item The flow speed applied in our experiments is $v = 2.5$ $\mu$m/s.
\item In our experiments, cells are tracked across the entire chamber with width $L = 1$ mm, excluding the microfluidic pillars that span $d = 100$ $\mu$m on either side. Therefore, the average and standard deviation of $G$ across the chamber are computed as
\begin{align}
\bar{G} &= \frac{1}{L-2d}\int_d^{L-d}dx\ G(x), \\
\delta G &= \sqrt{\left(\frac{1}{L-2d}\int_d^{L-d}dx\ G^2(x)\right)-\bar{G}^2},
\end{align}
where $G(x) = g(x)\sqrt{a^5/c(x)}$, with \cite{Moon2023}
\begin{align}
c(x) &= (c_2-c_1)\left(\frac{e^{vx/D}-1}{e^{vL/D}-1}\right) + c_1, \\
g(x) &= \frac{dc}{dx} = \frac{v(c_2-c_1)e^{vx/D}}{D(e^{vL/D}-1)},
\end{align}
and $a=10$ $\mu$m. Here, $c_1$ and $c_2$ are the TGF-$\beta$ concentrations on the upstream and downstream sides of the chamber, respectively, and $D = 150$ $\mu$m$^2$/s is the TGF-$\beta$ diffusion coefficient.
\item{Without flow ($v=0$), we used $c_1 = 0$ and $c_2 = 5$ nM, and cells migrated up-gradient on average [Fig.\ \ref{newdata}(g), red]. With flow: (i) when there was no gradient ($c_1=c_2=0$), cells migrated upstream on average [Fig.\ \ref{newdata}(g), blue]; (ii) when the gradient was small ($c_1 = 2.5$ nM, $c_2 = 5$ nM), cells migrated upstream on average [Fig.\ \ref{newdata}(g), purple]; and (iii) when the gradient was large ($c_1 = 0$, $c_2 = 10$ nM), cells migrated up-gradient on average [Fig.\ \ref{newdata}(g), brown].}
\end{itemize}
For Fig.\ \ref{phase}(c):
\begin{itemize}
\item Typical cell surface concentrations of ephrin are on the order of $c_2 = 10^2$$-$$10^3$ $\mu$m$^{-2}$, and ephrin receptors have been measured to respond to these concentrations (see Fig.\ 3 in \cite{xu2011epha2}). We use this range for the error bars in Fig.\ \ref{phase}(c).
\item The diffusion constant of ephrin was measured at $D_2 = 2$ $\mu$m$^2$/s (Fig.\ 2A in \cite{xu2011epha2}).
\item For head-to-head cell-cell collisions, the EGF concentrations used in \cite{lin2015interplay} increase by $\Delta c = 3.3$ nM on top of backgrounds of $c = 4.1$ nM, $c = 8.3$ nM, and $c = 12.4$ nM (see Fig.\ 3c-e in \cite{lin2015interplay}). Only $f = 40\%$ of this increase occurs within the cell microchannel, over a distance of approximately $d = 250$ $\mu$m, and the profile is linear within the microchannel (see Fig.\ S1e of \cite{lin2015interplay}). Thus, $g = f\Delta c/d$. These three values of $G = g\sqrt{a^5/c}$ with $a = 10$ $\mu$m are used in Fig.\ \ref{phase}(c).
\item The majority of cells migrated up-gradient (outcome 1 in Fig.\ 3a-e of \cite{lin2015interplay}) when $c = 4.1$ nM and $c = 8.3$ nM (see right bars in Fig.\ 3c and d of \cite{lin2015interplay}), whereas the majority of cells migrated away from the other cell (outcomes 2 and 3 in Fig.\ 3a-e of \cite{lin2015interplay}) when $c = 12.4$ nM (see right bars in Fig.\ 3e of \cite{lin2015interplay}).
\end{itemize}

\section{Experimental methods}
\label{exp}

\subsection{Cells and reagents}
KIC is a murine pancreatic cancer cell line derived from a genetically engineered murine pancreatic adenocarcinoma, featuring a Kras mutation along with Ink4a locus (Ink4a/ArfL/L) deletion \cite{sempere2011novel, varennes2019physical, whipple2012krasg12d}. In response to transforming growth factor-$\beta$ (TGF-$\beta$) stimulation, KIC cells exhibit a mesenchymal phenotype, enhanced invasive capacity, and directed migration \cite{moon2021signal, Moon2023, roussos2011chemotaxis}. KIC was cultured in RPMI 1640 supplemented with 1\% penicillin/streptomycin (P/S) and 5\% fetal bovine serum (FBS). Cells were routinely harvested using 0.05\% trypsin and 0.53 mM EDTA (Life Technologies, CA, USA) upon reaching approximately 80\% confluency in 25 cm$^2$ T-flasks. They were incubated at 37$^\circ$C with 5\% CO$_2$. The harvested cells were either used for experiments or sub-cultured, ensuring they remained below the 19th passage.

\subsection{Microfluidic platform}
The microfluidic device consists of two source channels for chemical gradients and one central channel for culturing cells with 100 $\mu$m thickness \cite{varennes2019physical, moon2021signal, Moon2023}. As shown in Fig.\ \ref{newdata}(a), source 1 and source 2 channels were connected to the reservoir to perfuse the medium into the central channel. The central channel width is 1 mm, which is filled with 2 mg/ml collagen matrix with cells. The concentration of TGF-$\beta$ (Human TGF-$\beta$1, Thermo Fisher) was controlled at source 1 and source 2 channels to generate chemical gradients in the central channel. To produce a 5 nM/mm TGF-$\beta$ gradient, the source 2 channel was filled with $c_2 = 5$ nM TGF-$\beta$, while the source 1 channel was filled with cell culture medium without TGF-$\beta$ ($c_1 = 0$). To create a small TGF-$\beta$ gradient with flow, the source 2 channel was filled with $c_2 = 5$ nM TGF-$\beta$, and the source 1 channel was filled with $c_1 = 2.5$ nM TGF-$\beta$. To create a large TGF-$\beta$ gradient with flow, the source 2 channel was filled with $c_2 = 10$ nM TGF-$\beta$, and the source 1 channel was filled with cell culture medium without TGF-$\beta$ ($c_1 = 0$). The initial cell density was maintained at $7\times10^5$ cells per ml across all groups. After loading the cells into the microfluidic chip, the cells were cultured in the regular medium for 24 hours. Subsequently, they were exposed to the engineered signaling environment under experimental conditions. To control the flow velocity at approximately 2.5 $\mu$m/s, we applied a pressure difference between two source channels \cite{Moon2023}.

\subsection{Characterization of cell migration}
Live-cell time-lapse imaging was performed using an inverted microscope with Zeiss incubation system (Zeiss AXIO Observer 7, Germany) at 37$^\circ$C with 5\% CO$_2$. KIC cells were imaged every 15 minutes for 3 hours and cell trajectories were analyzed using ImageJ. In gathering cell trajectories, we excluded trajectories of dividing cells and stationary cells. Dividing cells were excluded to prevent effects on cell polarity, while stationary cells were excluded to avoid underestimating cell movement characteristics. Directed cell migration is defined by motility and directional accuracy. The direction of cell trajectories was evaluated relative to environmental signals. Directional accuracy was quantified using the directional accuracy index, DAI $= \cos\theta$, where $\theta$ represents the angle between a trajectory's net displacement and the direction of the environmental cue [Fig.\ \ref{newdata}(b)]. The displacement is defined by a straight line connecting the trajectory's initial and final points. The DAI ranges from $-1$ to $1$. A DAI of 1 signifies that the cell moves precisely in the direction of the environmental signal, while a DAI of 0 indicates random motion. Conversely, a DAI of $-1$ means the cell moves in the exact opposite direction of the signal. A more detailed explanation of DAI can be found in previous studies \cite{varennes2019physical, Moon2023}.

\subsection{Analysis of DAI distributions}
The distributions in Fig.\ \ref{newdata}(g) were analyzed using the one-sample Wilcoxon ($p_{\rm W}$) test and the nonparametric Mann-Whitney test ($p_{\rm MW}$) for statistical comparison. Both the TGF-$\beta$-only condition ($p_{\rm W} < 0.0001$, $p_{\rm MW} < 0.0001$) and the flow-only condition ($p_{\rm W} < 0.01$, $p_{\rm MW} = 0.0862$) are significantly different from the control. As in \cite{Moon2023}, the small- and large-gradient conditions with flow are significantly different from each other ($p_{\rm W} = 0.0247$, $p_{\rm MW} = 0.0145$).

%\bibliography{refs.bib}


\end{document}